# Enhanced Critical parameters of *nano*-Carbon doped $MgB_2$ Superconductor

Monika Mudgel[1, 3], L. S. Sharath Chandra[2], V. Ganesan[2], G. L. Bhalla[3], H. Kishan[1] and V. P. S. Awana[1,*]

[1] National Physical Laboratory, Dr K.S. Krishnan Road, New Delhi-110012, India

[2]UGC-DAE Consortium for Scientific Research, University Campus, Khandwa Road, Indore-452017, India

[3]Deptartment of Physics and Astrophysics, University of Delhi, New Delhi-110007, India

## Abstract

The high field magnetization and magneto transport measurements are carried out to determine the critical superconducting parameters of $MgB_{2-x}C_x$ system. The synthesized samples are pure phase and the lattice parameters evaluation is carried out using the Rietveld refinement. The *R*-*T*(*H*) measurements are done up to a field of 140 kOe. The upper critical field values, $H_{c2}$ are obtained from this data based upon the criterion of 90% of normal resistivity i.e. $H_{c2}$=*H* at which $\rho$=90%$\rho_N$; where $\rho_N$ is the normal resistivity i.e., resistivity at about 40 K in our case. The Werthamer-Helfand-Hohenberg (*WHH*) prediction of $H_c(0)$ underestimates the critical field value even below than the field up to which measurement is carried out. After this the model, the Ginzburg Landau theory (*GL* equation) is applied to the *R*-*T*(*H*) data which not only calculates the $H_{c2}(0)$ value but also determines the dependence of $H_{c2}$ on temperature in the low temperature high field region. The estimated $H_c(0)$=157.2 kOe for pure $MgB_2$ is profoundly enhanced to 297.5 kOe for the x=0.15 sample in $MgB_{2-x}C_x$ series. Magnetization measurements are done up to 120 kOe at different temperatures and the other parameters like irreversibility field, $H_{irr}$ and critical current density $J_c(H)$ are also calculated. The *nano* carbon doping results in substantial enhancement of critical parameters like $H_{c2}$, $H_{irr}$ and $J_c(H)$ in comparison to the pure $MgB_2$ sample.

*Corresponding Author:

Dr. V.P.S. Awana

Fax No. 0091-11-25626938: Phone no. 0091-11-25748709

e-mail-awana@mail.nplindia.ernet.in: www.freewebs.com/vpsawana/

## Introduction

In the early years of discovery of renowned $MgB_2$ superconductor, it attracted the huge interest of scientific community due to it's simple chemical composition, crystal structure and highest $T_c$ among the intermetallic non-cuprate compounds [1-3]. The compound was studied extensively both by experimental and theoretical aspects by various groups. Soon, the typical and peculiar properties of $MgB_2$ came into picture like the two band nature having double band gap and the unusual Fermi surface topology [4, 5]. Various groups studied the band structure unfolding the mystery of different nature of Fermi surfaces for different [3-4, 6-7] bands. $MgB_2$ has two bands namely $\sigma$ and $\pi$. The Fermi surface due to $\sigma$ band has cylindrical sheets while possess tubular networks due to $\pi$ band. After all these studies on structural, electronic and band related properties of $MgB_2$ [6-9], the next step is to determine the effect of this two band nature on the critical properties of $MgB_2$ to estimate it's practical value. The effect of two band nature on critical parameters like upper critical field, $H_{c2}$ is needed to be probed. The $H_{c2}$ increases linearly near $T_c$ with decreasing temperature but it's behavior changes in the low temperature high field region. A sharp jump is predicted by theoretical and experimental reports near $T$=0 K in the $H_{c2}$ vs $T$ line [10-12]. That's why the exact $H_{c2}$ (0) value is much higher than it seems to be through normal extrapolation of data. The Werthamer-Helfand-Hohenberg (*WHH*) formula determines $H_{c2}$ (0) value on the basis of slope of $H_{c2}$ vs $T$ line at $T=T_c$. But since the slope is varying with the temperature considerably, it results in the wrong estimation of $H_{c2}$. After that Ginzburg-Landau theory is used for the calculation of $H_{c2}$ (0). The experimental data fits very well with the *GL* equation and the value of $H_{c2}$ (0) is found to be much higher than the *WHH* formula.

The critical properties of $MgB_2$ can be enhanced by *nano*-particle doping [13-15]. So, along with $MgB_2$, the *nano* carbon doped samples are also taken into consideration. The critical parameters like $H_{c2}$, $H_{irr}$ and $J_c$ enhances significantly by *nano*-carbon substitution at Boron site. The values of critical parameters obtained are either competitive or superior than those obtained earlier. Critical current density, $J_c$ increases by more than an order with *nano* carbon doping as estimated from magnetization plots. The substitution at boron site is more effective than *nano* particle additions in $MgB_2$ matrix. That's why the present results are superior than those of *nano*-SiC doping [14] for the optimum content. Actually in the $MgB_{2-x}C_x$ system, substitution of carbon at boon site results in intrinsic flux pinning along with the extrinsic pinning by excess carbon, not going at boron site but present at grain boundary. So, It enhances the critical parameters both ways and results in superior results than through other dopants [13-15]. Substantial increment is noticed in the $H_{c2}$ (0) value for the *nano* carbon doped samples as compared to pristine $MgB_2$ on applying suitable theoretical model. The quantitative description is given in the Results and discussion section and is also compared with the literature. Thus, hereby, we revisit our earlier studied $MgB_{2-x}C_x$ series [16] with high field Magneto-transport study up to 140 kOe applied field in this article. The transition temperature is still 5.80 K for the $MgB_2$ sample while the same is 12.80 and 11.30 K for the x=0.10 and x=0.15 sample at 140 kOe. So, the $H_{c2}$ (0) can-not be obtained experimentally. To determine $H_{c2}$ (0) value, we applied different theoretical models like *WHH* formula and Ginzburg Landau theory. Magnetization measurements confirm the enhanced critical parameters for carbon doped samples in comparison to pure $MgB_2$ sample.

## Experimental

The Polycrystalline $MgB_{2-x}C_x$ samples were synthesized by solid-state reaction route in the argon environment. The detailed procedure of synthesis of samples is given in Ref [16]. X-ray diffraction pattern were taken on Rigaku-Miniflex-Ultima Desktop diffractometer. Rietveld refinement was carried out using the software Fullproof-2007. Resistivity measurements were made on bar shaped samples using four-probe technique under the constant applied field on *Quantum Design PPMS*. Magnetization measurements were also carried out on *Quantum Design PPMS* equipped with *VSM* attachment.

## Results and Discussions

The X-ray diffraction patterns for the pristine and some of the *nano* carbon doped samples is shown in Fig. 1(a). Phase purity is checked by Rietveld refinement; all Bragg peaks are obtained at exact position with appropriate intensity. A small intensity extra phase MgO peak is also noticed in the pattern of $MgB_2$, which is marked by the symbol * in the figure. The *nano* carbon doped samples have the similar patterns with the shifted peaks according to the changed lattice parameters. The (100) peak shifts towards higher angle side, shown in the inset of Fig. 1(a), indicates towards the continuous decrease in *a* parameter. Rietveld refinement is done on all the samples and the so obtained lattice parameters are tabulated in Table I. '*a*' parameter decreases continuously as expected with the increase in *nano* carbon content in $MgB_{2-x}C_x$ samples, while *c* parameter does not change much. For pure $MgB_2$ sample, the lattice parameter *a* is found to be 3.0857(8) Å and the same decreases to 3.0678(20) Å for the highest *nano* carbon doped sample. The variation of lattice parameters, *c/a* value and cell volume with the increasing *nano* carbon content is shown in Fig.1 (b). Error bars for the lattice parameters '*a*' and '*c*' are also drawn as obtained from Rietveld Refinement. Cell volume and lattice parameter '*a*', both decrease with increase in x (*nano* carbon content in $MgB_{2-x}C_x$) while *c/a* value increases with the increasing *nano* carbon amount because of decreasing *a* parameter and almost constant *c* value. The continuous monotonic change in lattice parameters confirm the substitution of *nano* carbon at boron site in $MgB_2$ matrix but still the exact amount of *nano* carbon substituted at boron site is not known. The exact carbon content in $Mg(B_{1-y}C_y)_2$ is evaluated indirectly using the equation y = 7.5 × $\Delta c/a$, where $\Delta c/a$ is the change in *c*/*a* value as compared to the pure sample and y is the exact content by atomic wt. % of *nano* carbon substituted at the boron site[17-19]. The exact value calculated in this way is found to be quite less than expected. The net maximum substitution level is just 6% by atomic weight while the samples were prepared up to 10% by atomic weight. The x=0.2 i.e., $MgB_{1.80}C_{0.20}$ or $Mg(B_{0.90}C_{0.10})_2$ corresponds $Mg(B_{0.94}C_{0.06})_2$ or 6% by atomic weight, instead of nominal 10wt%. The remaining *nano* carbon stays at the grain boundary or at interstitial site and acts as a pinning centre and hence helps in enhancing the $H_{c2}$, $H_{irr}$ and $J_c(H)$ values. This is called the extrinsic pinning. The net carbon, which exactly goes at the boron site creates disorder in the sigma band and cause intrinsic pinning to enhance the critical parameters. So, Substitution by carbon at boron site causes extrinsic/intrinsic pinning through additions/substitution and is enhancing the superconducting performance of $MgB_2$ both ways. The variation of exact carbon content, y with the experimentally doped *nano* carbon content by atomic weight % is also plotted in the bottom layer of Fig.1 (b). The observed variation in the lattice parameters is in confirmation with the earlier reports, pertaining to carbon doping in $MgB_2$ [17, 20]

Fig. 2(a), 2(b) and 2(c) depict the variation of Resistivity with temperature in the transition zone at different field values varying from 0 to 140 kOe for the undoped, x=0.10 and 0.20 sample respectively. Here, we note that the transition is very sharp at zero field for all the samples but the transition width increases with the increase in field value. At low fields, behavior of pure sample is better than that of doped samples. The transition temperature $T_c$ ($\rho$=0) is 37.75 K for pure $MgB_2$ while it decreases with the boron site *nano* carbon substitution to 35.95K and 34.95K for x=0.10 and 0.20 samples respectively at zero field value. With increment in applied field, resistance curves shift towards lower temperature side both for doped & undoped samples but we can clearly see that relative shift is much lesser in case of doped sample curves than the pure one. Transition temperature for pure $MgB_2$ sample is only 5.80 K under 140 kOe field, while is increased to 12.80 K and 11.30 K for x=0.10 & x=0.15 samples respectively. So, addition of *nano* carbon clearly improves the superconducting performance of bulk $MgB_2$ sample at elevated fields. It simply implies that the critical field increases with the *nano* carbon doping in $MgB_2$. The transition temperatures $T_c$(ρ=0) for all the synthesized samples at fields varying from 0-140 kOe are given in Table II. Moreover, the normal state resistivity ($\rho_N$) also increases from 35 μΩ -cm for pure $MgB_2$ to about 140 μΩ -cm for x=0.10 and 0.20 samples [see Fig. 2(a), 2(b) &2(c)]. The increased value of normal state resistivity with *nano* carbon doping indicates towards the increased impurity scattering. The value of upper critical field especially $H_{c2}(0)$ is found to depend directly on $\rho_N$.[10] So, this observation is also in confirmation to the enhanced $H_{c2}$ for *nano* carbon doped samples. The variation of normalized resistivity ($\rho_T/\rho_{40}$) with temperature for un doped and some of the *nano* carbon doped samples is shown in Fig. 2(d). According to the definition of residual resistivity ratio, *RRR* value(=$\rho_{300}/\rho_{40}$) , the value of normalized resistivity($\rho_T/\rho_{40}$) at the end point of curves in Fig. 2(d) i. e. at 300 K directly corresponds to the *RRR* value for a particular sample. The *RRR* value is also plotted with the varying carbon content in the inset of Fig. 2(d). Pure sample is found to have highest value of *RRR* (=3.6) among the whole series of $MgB_{2-x}C_x$ samples. With increase in *nano* carbon content,

the *RRR* value has a monotonic decrease and the least value of 1.70 is obtained for the highest doped x=0.20 sample. It decreases very sharply in the beginning up to x=0.04 sample and after that rate of decrease in *RRR* value with respect to the increasing *nano* carbon content decreases. The *nano* carbon doping enhances the electron scattering in the doped sample and hence results in the decreased value of *RRR*. The above trend of change in *RRR* values of our samples is in confirmation with the literature [17, 18]

The critical field is determined for all the samples using the criterion that $H_{c2}=H$ at which $\rho=90\%\rho_N$ and $\rho_N$ is the normal resistivity or resistivity at about 40 K. The transition temperature with this criterion of $\rho=90\%\rho_N$ instead of $\rho=0$ are also determined for all the samples and are tabulated in Table III. The value of applied field in a column directly corresponds to the $H_{c2}$ value at the temperature given below in that column for corresponding samples. The variation of critical fields with temperature is shown in Fig. 3 for undoped as well as the *nano* carbon doped samples. At lower fields of less than 30 kOe, all samples have competing value of $H_{c2}$ but as the field increases, performance of *nano* carbon doped samples become far better than the undoped sample. As the carbon content increases, $H_{c2}$ also rises and the performance of x=0.08, 0.10 & 0.15 at higher fields is found to be competitive and best among this batch of samples. The other samples with 0.08>x>0.15 have slightly inferior performance but still it is quite better than the pure sample. This is because for the samples with x<0.08, the optimum level of *nano* carbon substitution is not reached yet and for x>0.15, the *nano* carbon may not go at the boron site and remains at the grain boundary. This can also induce grain boundary pinning but after a limit agglomeration of *nano* carbon particles take place so that the size of agglomerated clusters no longer remain of the range of coherence length of $MgB_2$ and become unable to pin the vortices. At 18.5 K the critical field of $MgB_2$ is near about 100 kOe while the same is increased to 140 kOe for x=0.10 *nano* carbon doped sample and lies in the range 120-140 kOe for other *nano* carbon doped samples. But since the measurements are done only up to 140 kOe and the temperature is still 18.5K for the x=0.10 sample, it is not possible to find $H_{c2}$ at lower temperatures experimentally. So, some theoretical models are need to be applied to see the behavior of upper critical field at low temperatures.

The simplest model to determine the upper critical field value at zero K i.e. $H_{c2}(0)$ is the Werthamer-Helfand-Hohenberg (*WHH*) formulation.

According to *WHH* formula

$$H_{c2}(0) = 0.69 * T_c * (dH_{c2}/dT)_{\text{at } T=T_c} \quad (1)$$

For x=0.10 sample, $H_{c2}(0)$ is just equal to 95 kOe by above formula which is not at all acceptable because the critical field of 140 kOe is already achieved at a temperature of 18.5 K. So it is not possible that critical field decreases with decrease in temperature. So, hereby we discard this formula for our system because it underestimates the $H_{c2}$ (0) value. This is also discussed by X. Huang et al [21] that $H_{c2}$ (0) value calculated by *WHH* formula is lesser than the real value by a factor of 5 or 6.

Another model applied for $H_{c2}$ determination is Ginzburg-Landau theory. The *GL* equation [22] in two band superconductors like $MgB_2$ for temperature dependence of $H_{c2}$ is given by

$$H_{c2}(T) = (H_{c2}(0) * \theta^{1+\alpha}) / (1-(1+\alpha)\omega + l\omega^2 + m\omega^3) \quad (2)$$

Where $\theta = 1-T/T_c$ and $\omega = (1-\theta)*\theta^{1+\alpha}$

The fitting of $H_{c2}$ vs $T$ data is done according to Equation 2. Both experimental and fitted curves for $H_{c2}$ are shown in Fig. 4. The Fitted curves are in solid line while experimental data points are shown by symbol. The theoretical curve fits very well with the experimental data up to the limit we carry out the measurements. So, the $H_{c2}$ line is drawn theoretically according to Eq.2. From the fitting, we can clearly see that, initially the behavior of $H_{c2}$ with $T$ is linear near $T_c$ and extends up to a temperature of 10 K and after that it saturates in the range of 3-10 K. Below 3 K the $H_{c2}$ line have negative curvature. The $H_{c2}$ (0) for x=0.15 sample is found to be about 300 kOe while the same is just nearly 160 kOe for the pure $MgB_2$ sample. All the *nano* carbon doped samples have $H_{c2}$ (0) values higher than the undoped sample. So, *GL* theory also confirms the enhancement of $H_{c2}$ with carbon doping in $MgB_2$ and determines the $H_{c2}$ (0) value. The exact values of $H_{c2}$ (0) for all samples is written in the inset of Fig. 4. The $H_{c2}$ (0) value determined by us matches well with Askerzade et al[23] for the undoped sample and in addition we have applied the same on *nano* carbon doped samples and achieved a considerable high value of 300 kOe. The $H_{c2}$(0) values determined for the *nano* carbon doped samples are also in confirmation with other reports in which high field measurements by pulsed magnetic field are carried out [23].

There is one more model known as Gurevich theoretical model for two band superconductors [11]. It takes into account the impact of both bands on the critical parameters. If we would have applied this model, the $H_{c2}$(0) value had been obtained as high as 400 kOe [12,24] in case of bulk and 500 kOe in case of thin films [10,25]. This actually corresponds to the real situation in case of $MgB_2$ because the negative curvature in $H_{c2}$ line near $T$=0 K according to *GL* equation is not expected. So, this theory proves very good for high temperature roughly above 5K. But below 5 K, the Gurevich model seems to be the best choice. Such a high value of above 400 kOe is really appreciable which proves this material to be a merit candidate for practical applications against Nb based superconductors and *HTSC* materials.

The magnetization hysteresis loop i.e., magnetization vs applied field curves are shown for doped and undoped samples in both increasing and decreasing field directions at 5, 10 and 20 K in inset of Fig. 5. The *M-H* loop for pure sample closes much before than the doped sample at each temperature, which clearly demonstrates the enhanced value of irreversibility field ($H_{irr}$). At 5 K, the loop closes nearly at about 80 kOe for the pure sample but is still open at 137 kOe for the *nano* carbon doped x=0.08 sample. All doped samples have better performance than the undoped samples. To have a clear idea, $H_{irr}$ (irreversibility field) are estimated for all samples at 5, 10 and 20 K from their respective magnetization loops. $H_{irr}$ is taken as the applied field value at which magnetization loop almost closes with a criterion of giving critical current density value of the order of $10^2$ A/cm$^2$. For pristine sample, the $H_{irr}$ values are 45, 74 & 80 kOe at 20,10 & 5K respectively, whereas it is increased to 63, 110 & 137 kOe for the x=0.08 sample at the same temperatures. These values are slightly higher than those reported earlier by Solatanian et al [26]. The increased values of $H_{irr}$ confirm the flux pinning by added *nano* carbon particles.

The critical current density is calculated from the magnetization hysteresis loops using Bean's Critical Model. The variation of $J_c$ with applied fields is shown in Fig. 5 for doped & undoped samples at 10K. All samples have $J_c$ of the order of more than $10^5$ A/cm$^2$ at low field values. As field increases, $J_c$ values decrease very rapidly for the pure sample and becomes of the order of $10^2$ A/cm$^2$ at a field of 60 kOe at 10K while it is still of the order of $10^4$ for the x=0.08 sample. Quantitatively, $J_c$ is about $1.04 \times 10^4$ A/ cm$^2$ at 60 kOe and 10K for x=0.08 *nano*-carbon doped sample, where as it is $5.4 \times 10^2$ A/cm$^2$ for pure sample at same field and temperature values. More specifically, $J_c$ of this sample is 21 times higher than the pure sample at 60 kOe & 10K. The critical current density value is enhanced similarly at other temperatures also (say 5 and 20 K) in the case of *nano* carbon doped samples. The ensuing pinning plots and the $J_c(H)$ performance of all samples at various temperatures are shown in ref. 16 by some of us. The observed values of $H_{c2}$, $H_{irr}$ and $J_c(H)$ are competitive or slightly better than those being reported yet [27-30].

## Conclusion

The *nano* carbon doped $MgB_{2-x}C_x$ system is studied for the enhanced critical parameters $H_{irr}$, $J_c(H)$ and especially the upper critical field $H_{c2}$. Theoretical models are applied on temperature dependence of upper critical field in order to estimate the critical field at low temperatures. $H_{c2}$ (0) for all the carbon doped samples is found to be higher than the pure $MgB_2$ sample. The $H_{c2}$ (0) value for pure sample is just 157 kOe which got profoundly enhanced and the highest value of $H_{c2}$ (0) of about 300 kOe is achieved for x=0.15 sample. The $H_{c2}$ (0) of about 400 kOe is expected by applying the new two band Gurevich model on this system. Not even the upper critical field but the other parameters like $H_{irr}$ and $J_c(H)$ are also improved significantly for the carbon doped samples.

## Acknowledgement

The authors from *NPL* would like to thank Dr. Vikram Kumar (*DNPL*) for his great interest in present work. Monika Mudgel would like to thank the *CSIR* for the award of Junior Research Fellowship to pursue her *Ph. D* degree.

References

1. J. Nagamatsu, N. Nakagawa, T. Muranaka, Y. Zenitani and J.Akimitsu, Nature **410,** 63 (2001).

2. S. L. Budko, J. Laperot, C. Petrovic, C. E. Cunningham, N. Anderson and P. C. Canfield, Phys. Rev. Lett. **86**, 1877 (2001)
3. J. Kortus, I. Mazin, K. D. Belashchenko, V. P. Antropovz and L. L. Boycry, Phys. Rev. Lett. **86**, 20 (2001)
4. J. M. An and W. E. Pickett, Phys. Rev. Lett. **86**, 4366 (2001)
5. H. Kotegawa, K. Ishida, Y. Kitaoka, T. Muranaka and J. Akimitsu, Phys. Rev. Lett. **87**, 127001 (2001)
6. G. Satta, G. Profeta, F.Bernardini, A. Continenza and S.Massida, Phys. Rev. B **64** 104507 (2001)
7. K. D. Beleschenko, M. Van Schilfgaarde and V. P. Antroprov, Phys. Rev. B **64** 092503 (2001)
8. J.D. Jorgensen, D.G. Hinks, and S. Short, Phys. Rev. B, **63** 224522 (2001)
9. S. Margadonna, T. Muranaka, K. Prassides, I. Maurin, K. Brigatti, R. M. Ibberson, M. Irai, M. Takata and J. Akimitsu, J. Phys. Cond. Matter **13**, L795 (2001)
10. V. Ferrando, P. Manfrinetti, D. Marre, M. Putti, I. Sheikin, C. Tarantini and C. Fedeghini, Phys. Rev. B **68**, 094517 (2003)
11. A. Gurevich, Phys. Rev. B **67**, 184515 (2003)
12. X. Huan, W. Mickelson, B. C. Regan and A. Zettl, Solid state Communications **136**, 278-282 (2005)

13. J. H. Kim, S. X. Dou, M. S. A. Hossain, X. Xu, X. L. Wang, D. Q. Shi, T. Nakane and H. Kumakura, Supercond. Sci. & Tech. **20**, 715 (2007)

14. A. Vajpayee, V.P.S. Awana, G. L. Bhalla and H. Kishan, Nanotechnology **19**, 125708 (2008)

15. A. Vajpayee, V.P.S. Awana, H. Kishan, A.V. Narlikar, G.L. Bhalla, X.L. Wang, J. Appl. Phys. **103**, 07C0708 (2008)

16. M. Mudgel, V. P. S. Awana, G. L. Bhalla and H. Kishan, Solid State Communications **146**, 330 (2008)

17. M. Avdeev, J. D. Jorgensen, R. A.Ribeiro, S. L. Bud'ko and P. C. Canfield, Physica C **387**, 301 (2003)

18. A. Bharathi, S. J. Balaselvi, S. Kalavathi, G. L. N. Reddy, V. S. Sastry, Y. Hariharan and T. S. Radhakrishnan, Physica C **370**, 211 (2002)

19. W. K. Yeoh and S.X. Dou, Physica C **456**, 170 (2007)

20. T. Takenobu, T. Ito, D. H. Chi, K. Prassides and Y. Iwasa, Phys. Rev. B **64**, 134513 (2001)

21. X. Huang, W. Mickelson, B. C. Regan and A. Zettl, Solid State Communications **136** 278 (2005)

22. I. N. Askerzade, A. Gencer and N. Guclu, Supercond. Sci. Technol. **15** L13 (2002)

23. R. H. T. Wilke, S. L. Budko, P. C. Canfield and D. K. Finnemore, Phys. Rev. Letter **92** 217003 (2004)

24. S. Noguchi, A. Kuribayashi, T. Oba, H. Iriuda, Y. Harada, M. Yoshizawa and T. Ishida, Physica C **463** 216 (2007).

25. V. Braccini, A. Gurevich, J. E.Giencke, M. C. Jewell, C. B. Eom and D. C. Larbalestier, Phys. Rev. B **71** 012504 (2005)

26. S. Solatanian, J. Horvat, X. L. Wang, P. Munroc and S. X. Dou, Physica C **390** 185 (2003)

27. Z. H. Cheng, B. Shen, J. Sheng, S. Y. Zhang, T. Y. Zhao and H.W. Zhao, J. Appl Phys. **91** 7125 (2002)

28. W. K. Yeoh, J. H. Kim, J. Horvat, X. Xu, M. J. Qin, S. X. Dou, C. H. Jiang, T. Nakane, H. Kumakura; P. Munroe, Supercond. Sci. & Tech. **19** 596 (2006)

29. M. Herrmann, W. Habler, C. Mickel, W. Gruner, B. Holzapfel and L. Schultz, Supercond. Sci. & Tech. **20** 1108 (2007)

30. W. K. Yeoh, J. H. Kim, J. Horvat, X. Xu and S. X. Dou, Physica C, **460** 568 (2007)

**Table I :** Lattice parameters, c/a values and cell volume is categorized for $MgB_{2-x}C_x$ samples (x=0.0, 0.02, 0.04, 0.06, 0.08, 0.10, 0.15 & 0.20)

| Sample | Atomic wt % of Carbon | *a* (Å) | *c* (Å) | *Volume* (Å$^3$) | *c*/*a* | Actual wt% of carbon |
|---|---|---|---|---|---|---|
| $MgB_2$ | 0 | 3.0857(8) | 3.5230(8) | 29.15 | 1.142 | 0 |
| $MgB_{1.96}C_{0.04}$ | 2 | 3.0803(7) | 3.5250(8) | 29.0 | 1.144 | 1.73 |
| $MgB_{1.92}C_{0.08}$ | 4 | 3.0754(16) | 3.5275(16) | 28.89 | 1.147 | 3.75 |
| $MgB_{1.90}C_{0.10}$ | 5 | 3.0742(24) | 3.5287(24) | 28.88 | 1.148 | 4.5 |
| $MgB_{1.85}C_{0.15}$ | 7. 5 | 3.0692(19) | 3.5271(20) | 28.77 | 1.149 | 5.25 |
| $MgB_{1.80}C_{0.20}$ | 10 | 3.0678(20) | 3.5336(21) | 28.80 | 1.151 | 6.75 |

**Table II** : Transition temperature ($T_c$ at R=0) at different field values (0 to 14T) for $MgB_{2-x}C_x$ samples

| Sr. No. | x in $MgB_{2-x}C_x$ | $T_c$ $H$=0 kOe | $T_c$ $H$=10 kOe | $T_c$ $H$=30 kOe | $T_c$ $H$=50 kOe | $T_c$ $H$=70 kOe | $T_c$ $H$=90 kOe | $T_c$ 110 kOe | $T_c$ 130 kOe | $T_c$ 140 kOe |
|---|---|---|---|---|---|---|---|---|---|---|
| 1 | 0.0 | 37.75 | 34.10 | 29.03 | 24.55 | 20.55 | 16.54 | 12.30 | 8.06 | 5.80 |
| 2 | 0.04 | 36.79 | 33.60 | 29.04 | 25.05 | 22.06 | 18.55 | 15.55 | 12.30 | 10.55 |
| 3 | 0.08 | 36.19 | 32.80 | 27.80 | 24.04 | 20.29 | 17.30 | 14.04 | 11.30 | 10.05 |
| 4 | 0.10 | 35.95 | 32.80 | 28.30 | 24.79 | 21.55 | 19.04 | 15.80 | 13.80 | 12.80 |
| 5 | 0.15 | 35.19 | 32.55 | 27.79 | 23.80 | 20.30 | 17.54 | 15.05 | 12.54 | 11.30 |
| 6 | 0.20 | 34.69 | 31.55 | 26.82 | 22.81 | 19.30 | 16.04 | 13.30 | 10.79 | 9.29 |

**Table III :** Transition temperature ($T_c$ at R=90%$R_{40}$) to determine $H_{c2}$ at different field values (0 to 14T) for $MgB_{2-x}C_x$ samples

| Sr. No. | x in $MgB_{2-x}C_x$ | $T_c$ $H$=0 kOe | $T_c$ $H$=10 kOe | $T_c$ $H$=30 kOe | $T_c$ $H$=50 kOe | $T_c$ $H$=70 kOe | $T_c$ $H$=90 kOe | $T_c$ 110 kOe | $T_c$ 130 kOe | $T_c$ 140 kOe |
|---|---|---|---|---|---|---|---|---|---|---|
| 1 | 0.0 | 38.84 | 35.10 | 30.73 | 27.28 | 24.01 | 20.58 | 17.60 | 14.18 | 12.47 |
| 2 | 0.04 | 37.72 | 34.79 | 30.88 | 27.81 | 25.30 | 22.63 | 20.28 | 18.02 | 16.77 |
| 3 | 0.08 | 37.19 | 34.25 | 30.48 | 27.63 | 25.20 | 23.02 | 20.91 | 18.99 | 17.98 |
| 4 | 0.10 | 37.04 | 34.11 | 30.52 | 27.93 | 25.52 | 23.42 | 21.50 | 19.56 | 18.57 |
| 5 | 0.15 | 37.10 | 33.97 | 30.29 | 27.62 | 25.05 | 22.82 | 20.76 | 18.73 | 17.77 |
| 6 | 0.20 | 36.64 | 33.43 | 29.53 | 26.87 | 24.37 | 22.0 | 20.05 | 17.92 | 17.00 |

## Figure Captions

Figure 1(a). X-ray diffraction patterns for the $MgB_{2-x}C_x$ series (x=0.0, 0.04, 0.10, & 0.15).

Figure 1(b). Variation of lattice parameters, cell volume and exact carbon content for $MgB_{2-x}C_x$ series (x=0.0-0.20).

Figure 2. Resistivity vs temperature plot at different field values varying from 0-140 kOe for (a) pure $MgB_2$ (b) $MgB_{2-x}C_x$, x=0.10 (c) $MgB_{2-x}C_x$, x=0.20

Figure 2(d). Variation of normalized resistivity ($\rho_T/\rho_{40}$) with temperature is shown for $MgB_{2-x}C_x$ series (x=0.0, 0.04, 0.10, 0.15 & 0.20). The RRR values are plotted with the carbon content in the inset.

Figure 3. $H_{c2}$ vs temperature plots for $MgB_{2-x}C_x$(x=0.0-0.20) samples.

Figure 4 Theoretically fitted curves for $H_{c2}$ vs temperature plots for $MgB_{2-x}C_x$(x=0.0-0.20)

Figure 5 $J_c(H)$ plots for $MgB_{2-x}C_x$ samples (x=0.08, 0.10 & 0.20) along with pristine $MgB_2$ at 10K in the main panel while inset shows the magnetization loop $M(H)$ at 5, 10 & 20K for $MgB_{2-x}C_x$ samples (x=0.0, 0.08, 0.10 & 0.20) up to 120 kOe field

Fig. 1(a)

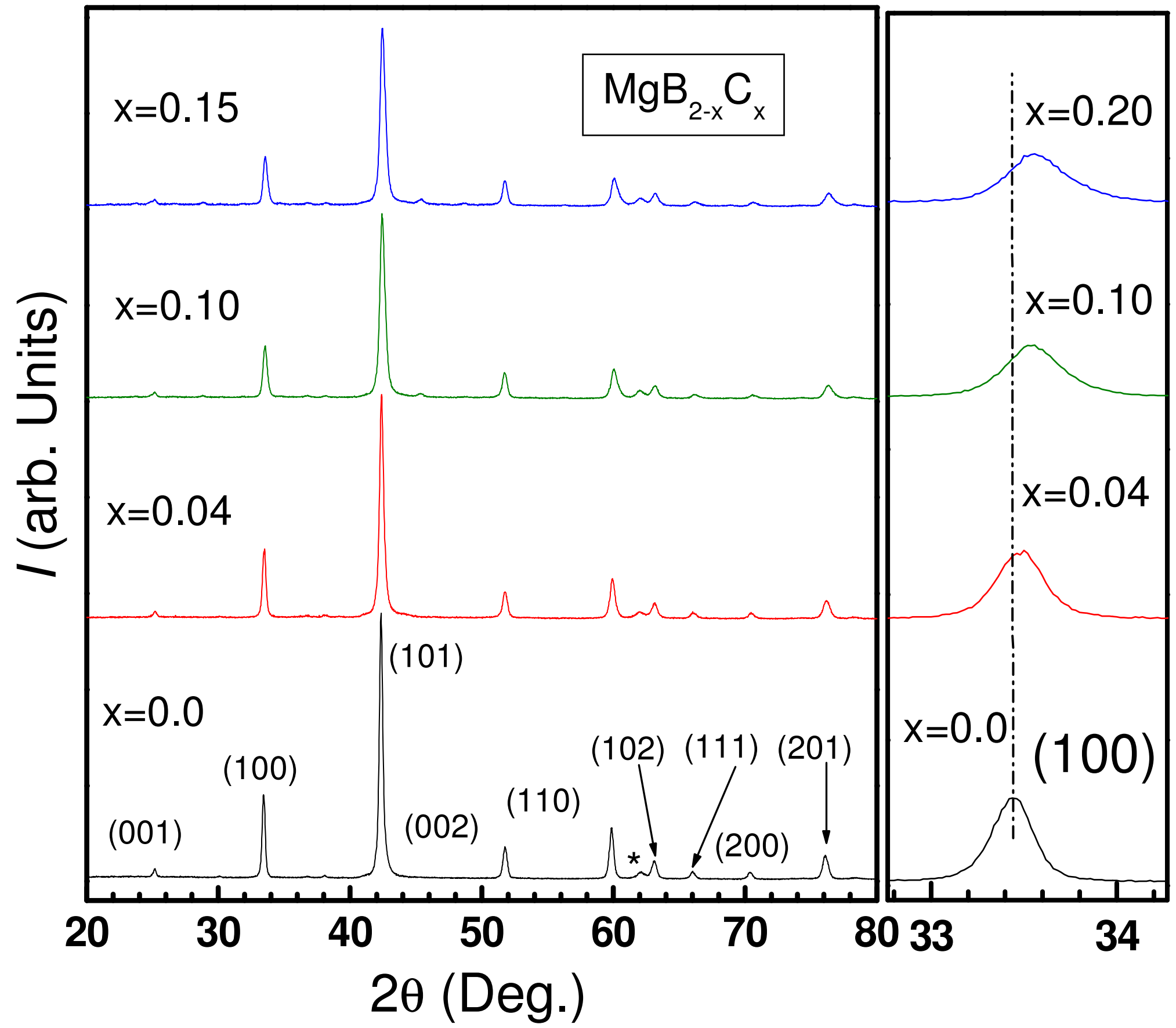

Fig. 1(b)

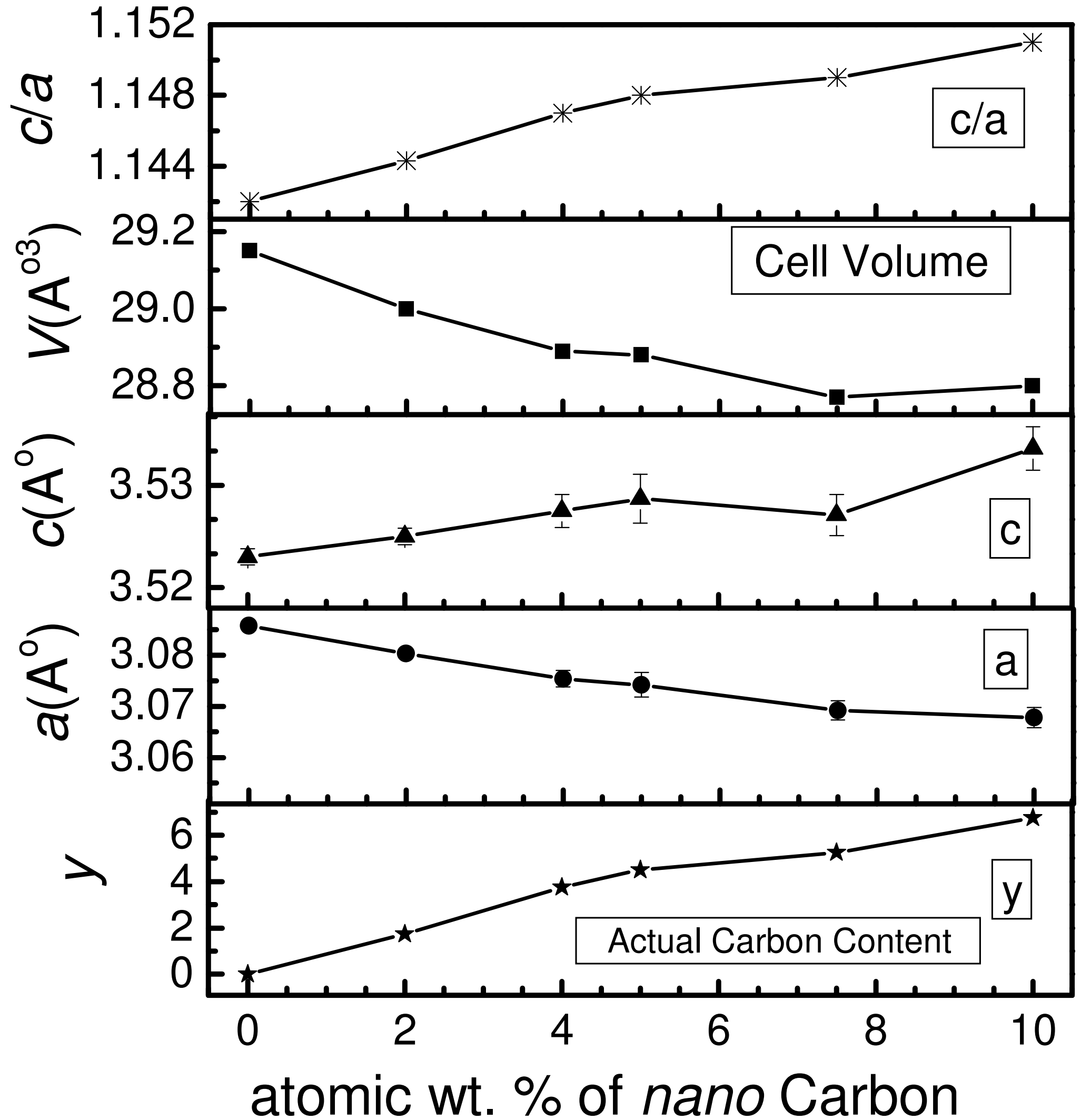

Fig. 2(a)

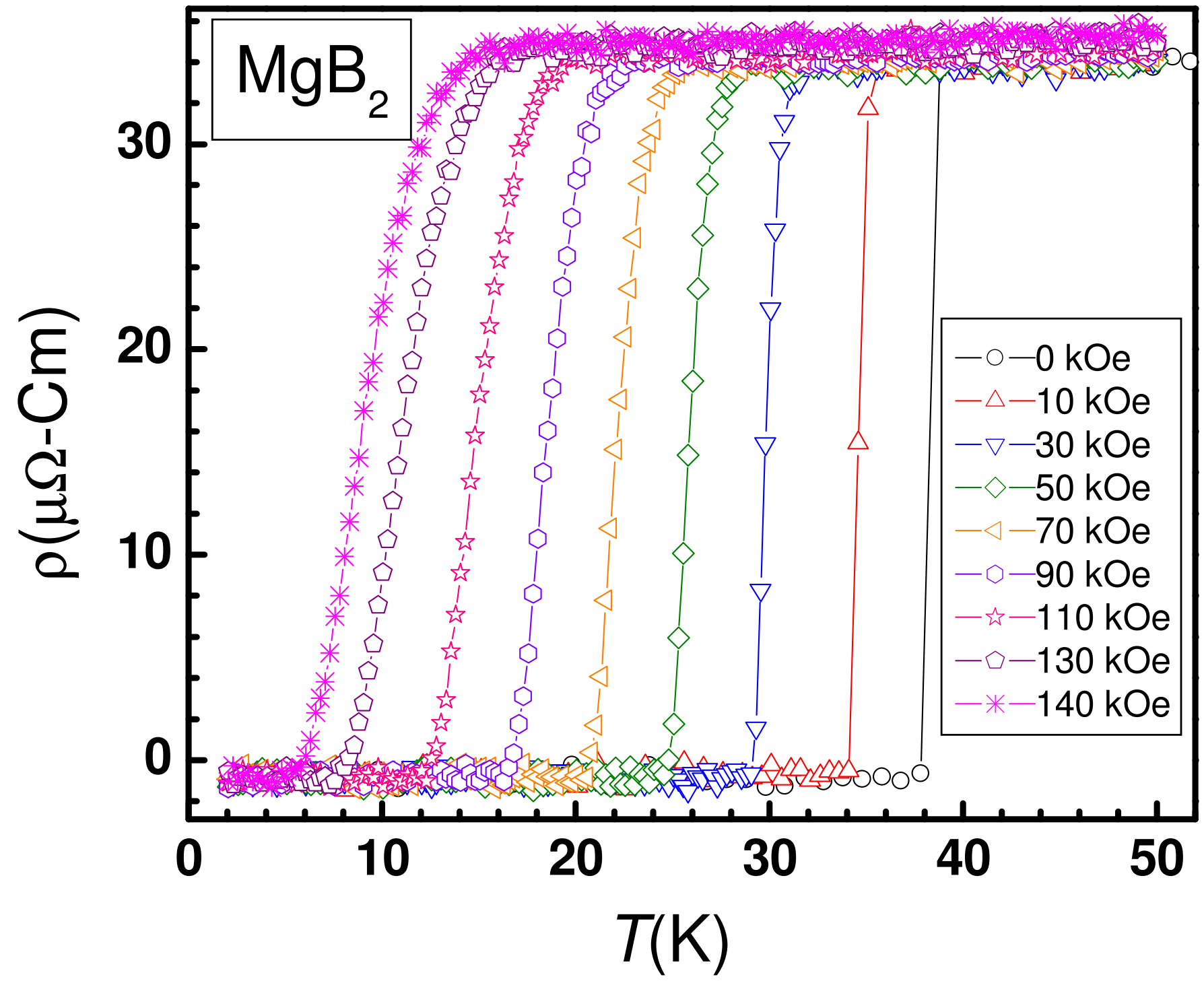

Fig. 2(b)

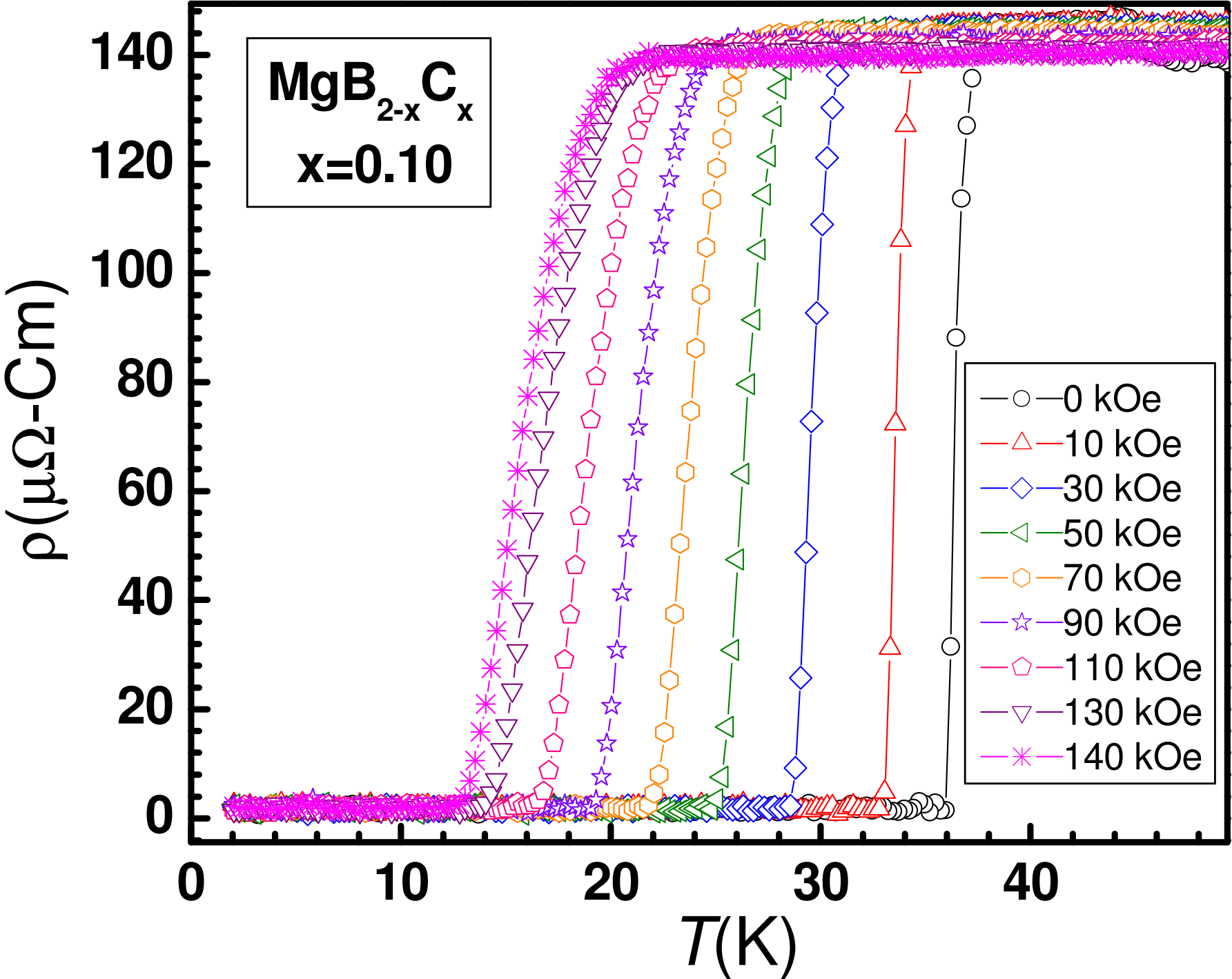

Fig. 2(c)

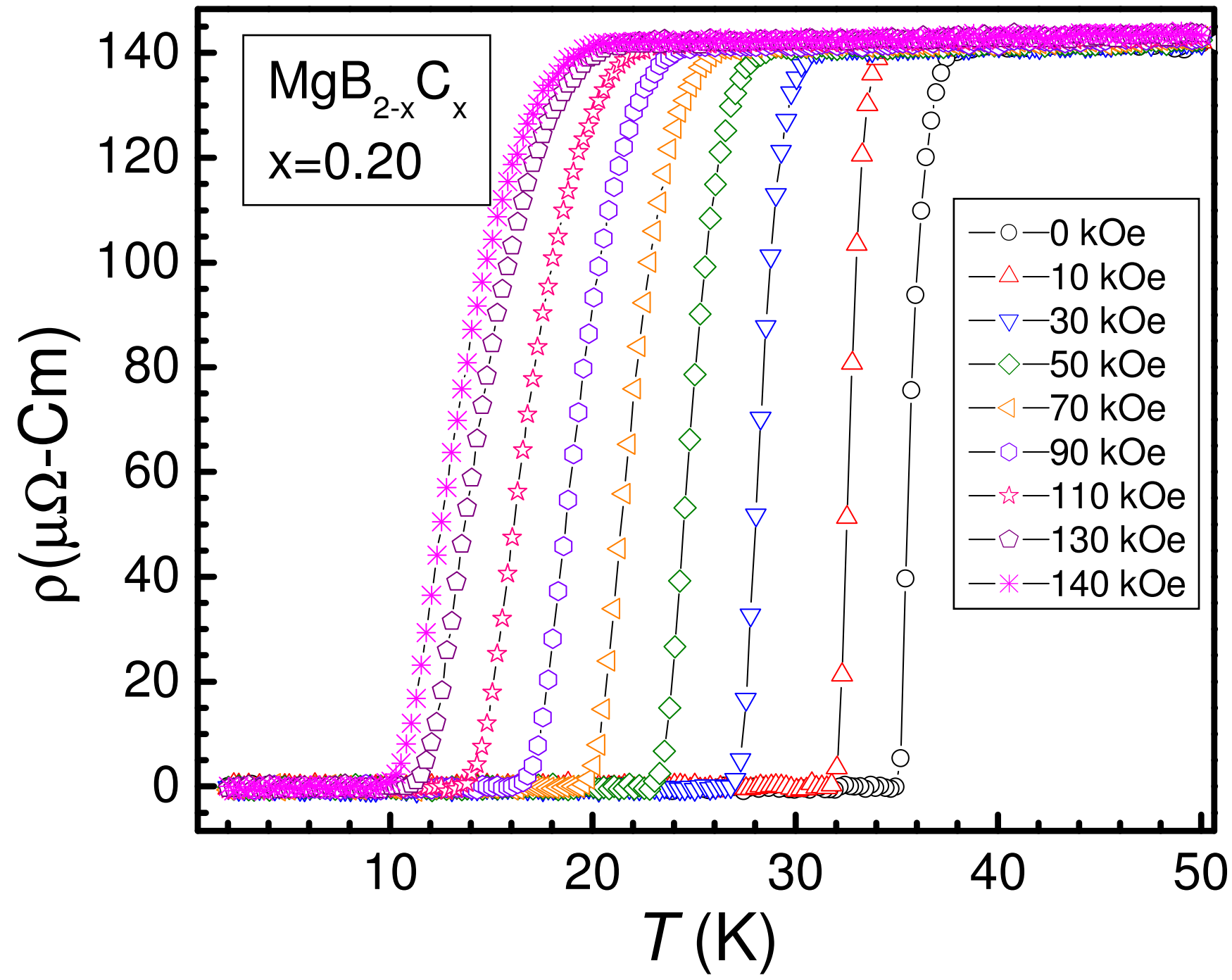

Fig. 2(d)

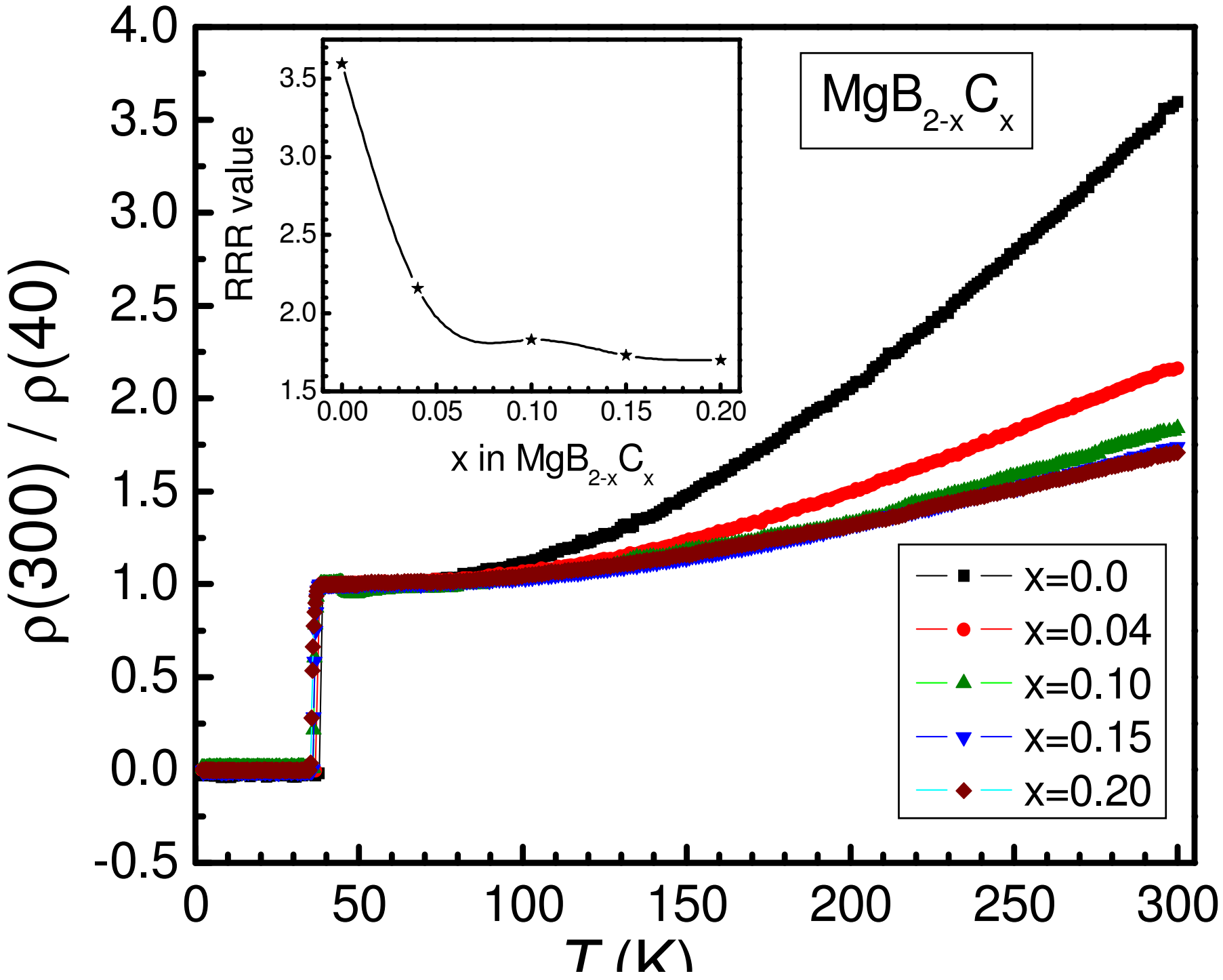

Fig. 3

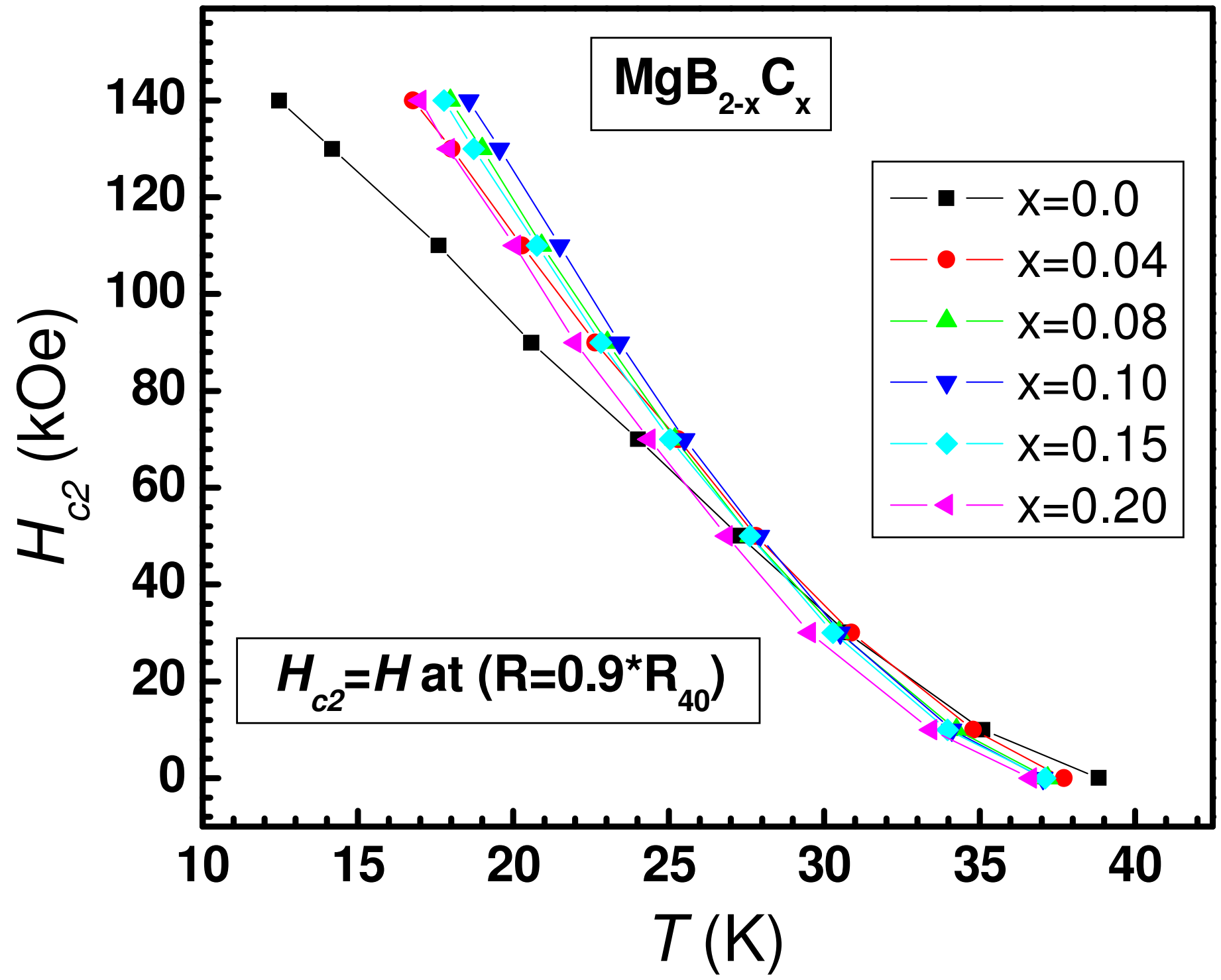

Fig.4

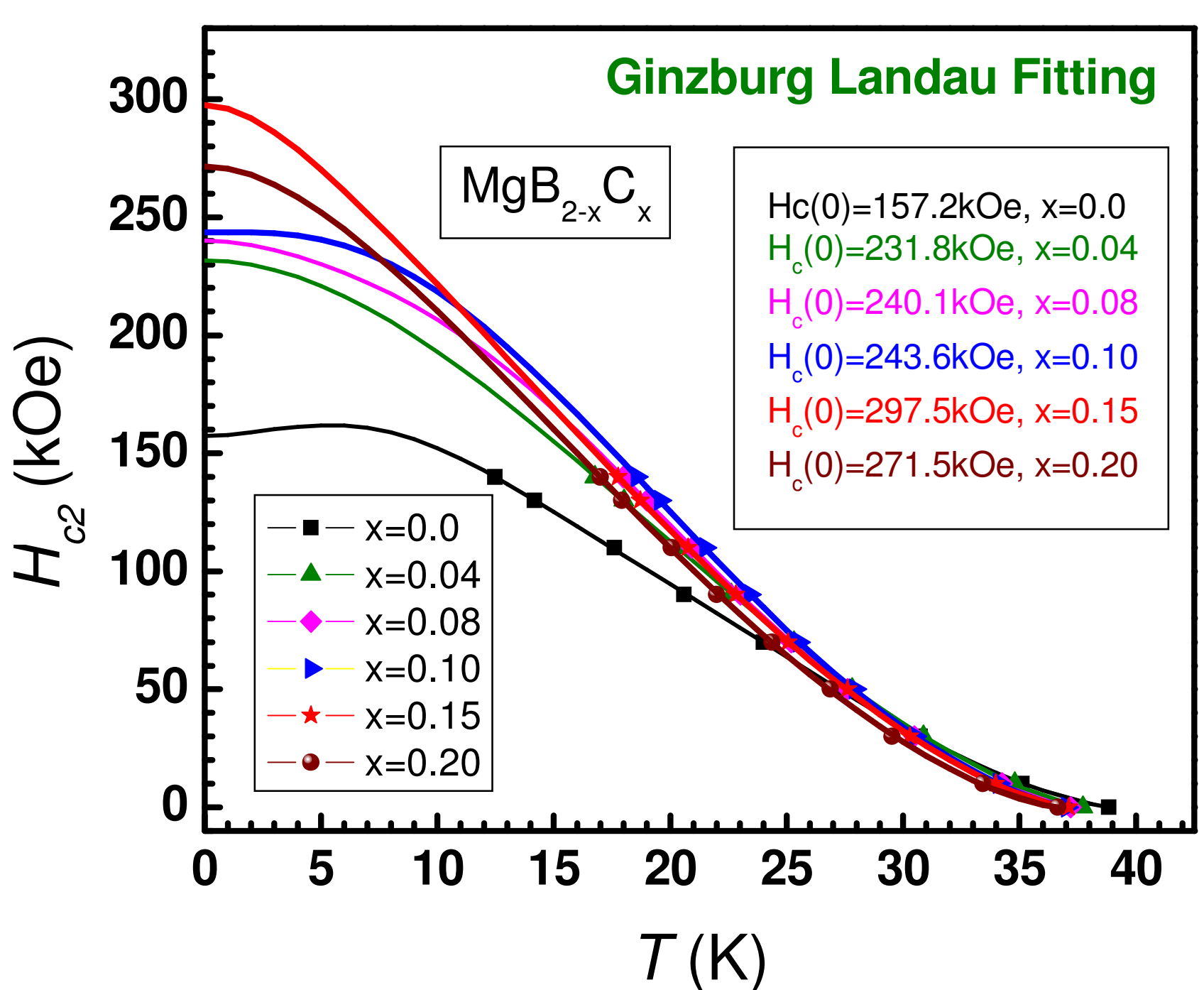

Fig. 5

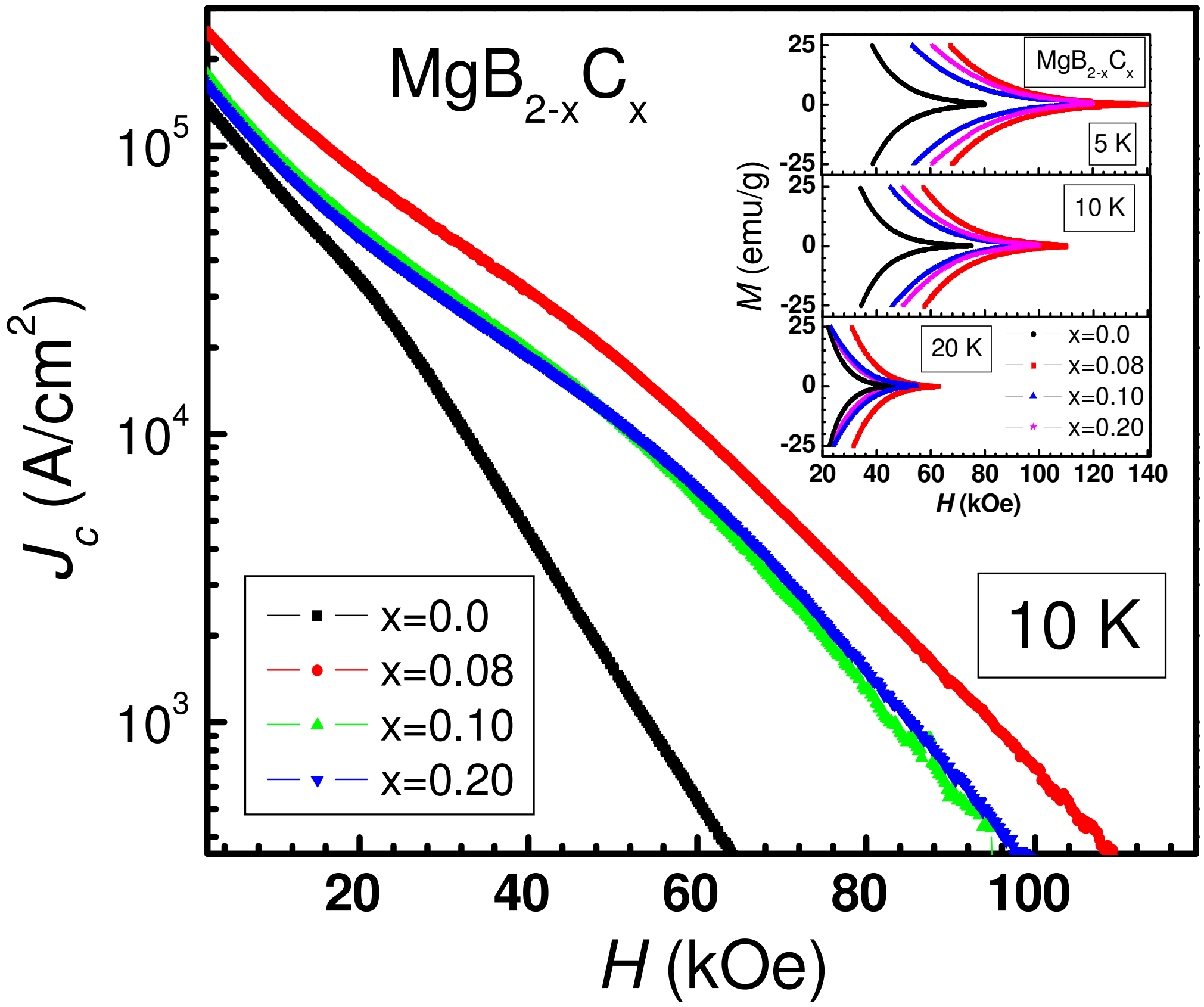